\documentclass[prl,twocolumn,aps,superscriptaddress,amsmath,amssymb]{revtex4-1}
\usepackage{graphicx,color,setspace}

\usepackage[normalem]{ulem}

\begin{document}

\title{Generating ultrastable glasses by homogenizing the local virial stress}
\author{Fabio Leoni}
\affiliation{Department of Physics, Sapienza University of Rome, P. le Aldo Moro 5, 00185 Rome, Italy}
\email{fabio.leoni@uniroma1.it}
\author{John Russo}
\affiliation{Department of Physics, Sapienza University of Rome, P. le Aldo Moro 5, 00185 Rome, Italy}
\author{Francesco Sciortino}
\affiliation{Department of Physics, Sapienza University of Rome, P. le Aldo Moro 5, 00185 Rome, Italy}
\author{Taiki Yanagishima}
\affiliation{Department of Physics, Graduate School of Science, Kyoto University, Kyoto, 606-8224, Japan}
\email{yanagishima.taiki.8y@kyoto-u.ac.jp}

\begin{abstract}
In recent years, the possibility of algorithmically preparing ultra-stable glasses (UG), i.e., states that lie very deep in the potential energy landscape, has considerably expanded our understanding of the glassy state. In this work, we report on a new protocol for ultrastable glass preparation that iteratively modifies the particle diameters to reduce local virial stress fluctuations. We apply the algorithm to an additive Lennard-Jones mixture and show that, compared to the states obtained via thermal annealing, virial homogenized glasses (VHG) are characterized by a considerable increase in both kinetic stability and the number of locally favored structures (icosahedra). We also consider the melting dynamics during heating ramps and show that it occurs via an accumulation of localized events. Our results highlight the connection between the thermodynamic and mechanical stability of ultra-stable glassy states.
\end{abstract}

\pacs{}
\maketitle

\date{\today}

Since pioneering work by Ediger and collaborators~\cite{Swallen2007}, the advent of ultra-stable glasses has opened new frontiers in the application and the understanding of glassy materials~\cite{Ediger2017,Rodriguez-Tinoco2022}. In experiment, ultra-stable glasses are commonly obtained from vapor deposition on a substrate at the right temperature and deposition rate, displaying an enhanced kinetic and thermodynamic stability with respect to annealed glasses. The same stability would require thousands of years of ageing when starting with annealed glasses \cite{Lyubimov2013}. Recent investigations have shown that the use of sample size effects \cite{monnier2021,dilisio2023} or `hyperaging' of systems with a glass transition just above room temperature\cite{zhao2022} may also be feasible strategies. The enhanced stability of ultra-stable glasses is reflected in many properties \cite{Dalal2013,Fullerton2017,Rodriguez-Tinoco2022,Sepulveda2014,Dalal2012,Whitaker2015}: these include resistance to permeation by inert gas \cite{Ediger2017}, resistance to crystallization \cite{Rodriguez-Tinoco2022,rodriguez2015}, the ability to modulate chemical reactivity \cite{Qiu2016}, and resistance to devitrification \cite{Ediger2017}. Promising applications of this technology include the realization of OLED displays \cite{Ediger2017}, the stabilization of amorphous pharmaceuticals \cite{Ediger2017}, the miniaturization of computing components~\cite{simon2020role,sil2021impact,jamison2015sio,rosenberg2000copper,nogami2022advanced}, and the realization of novel metallic glasses with improved mechanical properties~\cite{Luo2018,Ashby2006,wang2009}.

At the same time, the study of algorithms capable of generating glasses with enhanced properties is yielding new insights into the theoretical understanding of such states of matter.
Several computational works have shown that it is possible to simulate the process of vapor deposition to obtain states that are deeper in the potential energy landscape than those obtained by thermal annealing, even though the time-scales accessible to simulations prevents these glasses from displaying the same dramatic increase in stability observed in experiments~\cite{Lyubimov2013,reid2016age,Berthier2017}. Nevertheless, simulation studies of vapor deposition have been very successful in uncovering the physical mechanism which underlies both the formation of ultra-stable glasses, pointing towards the enhanced diffusivity of the moving surface as the key mechanism that promotes structural ordering~\cite{Ediger2017,Rodriguez-Tinoco2022,Berthier2017,leoni2023a}, and their melting behaviour~\cite{flenner2019front}.

In order to reach the same levels of ultrastability as those displayed in experiments, simulations have exploited preparation routes based on \emph{unphysical} moves. The most successful so far are \emph{swap} moves \cite{ninarello2017models,berthier2023,Kuchler2023}, where particles are free to swap positions, and which have proven to be effective in equilibrating size polydisperse mixtures well below the conventional (without swap) glass transition temperature~\cite{ninarello2017models,brito2018theory,ikeda2019effect}. Swap simulations close to the glass transition temperature have shown that relaxation is initially localized, and that successive events take place close to the original ones, which point towards a dynamic facilitation as the main ingredient for the dynamics at deeply supercooled conditions~\cite{scalliet2022thirty} and is responsible for the asymmetric wings observed in experimental relaxation spectra~\cite{guiselin2022microscopic}. The greatest strength of the swap algorithm can also be in some applications its biggest weakness, as its remarkable equilibration ability often leads to the crystallization of the system unless tailored size distributions are used~\cite{ninarello2017models} (as we verified to happen for our system described below). Other unphysical moves that promote the formation of ultrastable glasses have been introduced, such as the random bonding of monomers~\cite{ozawa2023creating,ozawa2023equilibrium}, random pinning of particles \cite{hocky2014,ozawa2015} or changing particle diameters~\cite{kapteijns2019fast,hagh2022transient,brito2018theory,gavazzoni2023testing,bolton2024ideal,fan2024ideal} to minimize the potential energy with some population constraint, which starts from an equilibrium configuration of a system and produce a ultra-stable (out of equilibrium) state of a modified system.

Recently, a new move which changes the particle sizes in a monodisperse repulsive system in order to \emph{homogenize} the local densities was shown to produce ultra-stable glasses that are only weakly polydisperse (size polydispersity of approximately 3\%)~\cite{Yanagishima2021}. These microscopically density `uniform' glass states did not show any aging within the time scales simulated, in contrast to the intermittent melting events seen in the quenched glass prepared with the same particle size distribution~\cite{Sanz2014,Yanagishima2017}. It was also found to show enhanced stability against crystallization when placed adjacent to a slab of crystal~\cite{Yanagishima2023}. The analysis of these ultra-stable states revealed a noteworthy homogenization in the number of load-bearing nearest neighbors for each particle, suggesting the potential for the formation of ultra-stable glasses through the homogenization of the mechanical environment of individual particles.

In this work, we aim to show that mechanical homogenization is not simply an emergent feature of ultrastable states, but may be applied as a direct means to achieve glass stabilization. While our previous efforts to homogenize local densities was easily applicable to monodisperse or weakly polydisperse systems~\cite{Yanagishima2021}, it could not be applied to multicomponent systems, where there is no physical reason for uniform density be stable. Not only is this a key roadblock to stabilizing e.g. widely studied binary systems, it shows that there may be better metrics to target. Specifically, we propose an algorithm that homogenizes the local virial stress fluctuations by incremental changes to individual particle sizes. The algorithm acts on the configurational part of the pressure tensor: as the distribution of virial stresses sharpens, this progressively reduces both the bulk pressure and the energy of the system without necessarily reaching an equilibrium state.

We will show that for an additive Lennard-Jones mixture our algorithm produces ultra-stable glasses with a considerable increase in the thermodynamic and kinetic stability of the system with respect to conventional glasses quantified by the potential energy and onset temperature. Moreover, we will show that the increased stability is associated with a drastic increase in the number of locally favoured structures, confirming the link between ultrastability and structural properties that was recently introduced in Ref.~\cite{leoni2023a}. Finally, we will analyze the glass melting dynamics, and show that it occurs through a cascade of localized events, similarly to what was recently discovered in deeply supercooled states~\cite{hocky2014,scalliet2022thirty,ruiz2023real,vila2023emergence,herrero2023}.

\section{Methods}\label{sec:methods}

\subsection*{Homogenization algorithm}

\begin{figure}[!t]
    \centering
    \includegraphics[width=1.0\linewidth]{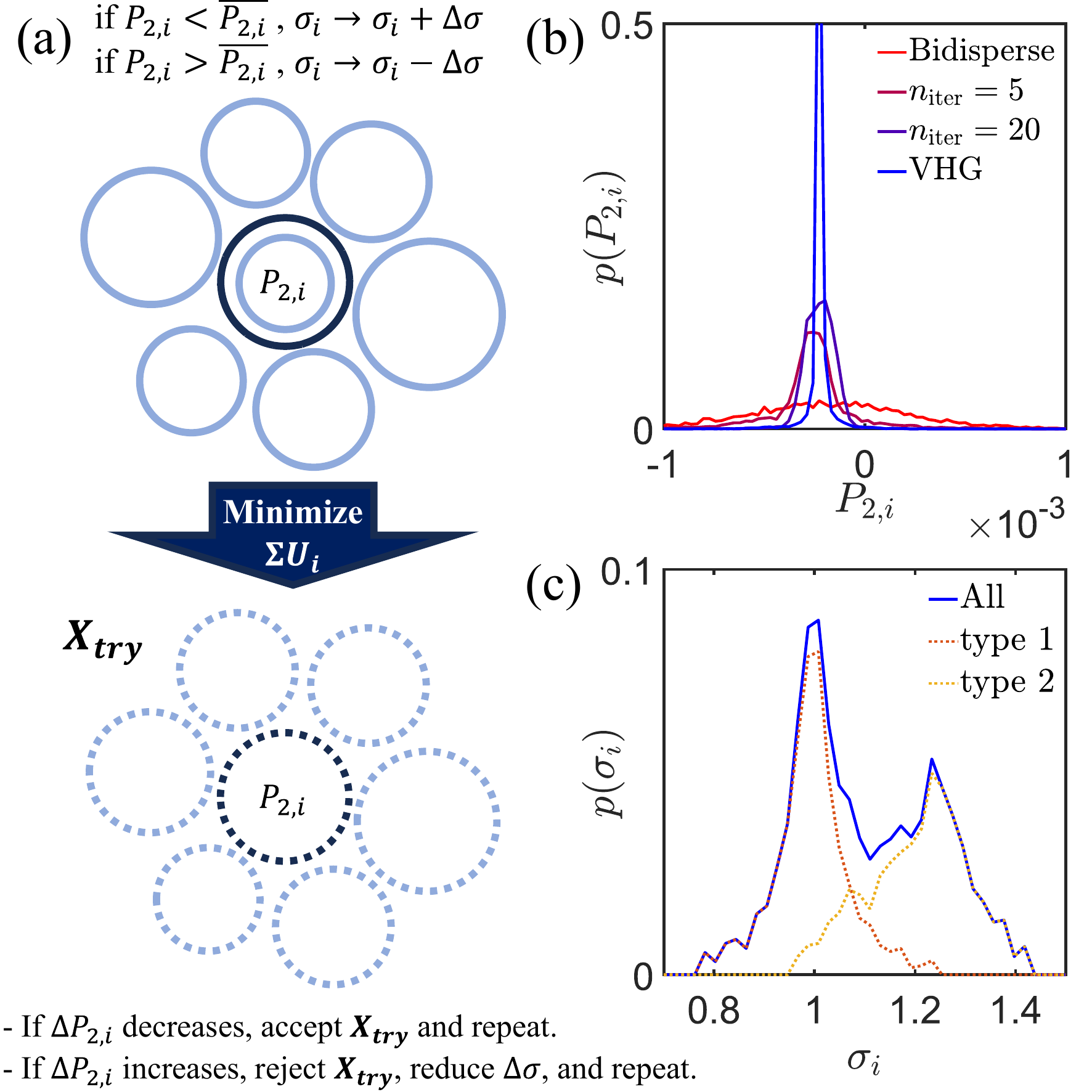}
    \caption{Creation of `virial homogenized' glasses. (a) Illustration of the iterative scheme used to make VHGs for a move where $dP_{2,i}/d\sigma_i > 0$. (b) The distribution of local virial pressures $P_{2,i}$ with iteration. (c) The size distribution $\sigma_i$ obtained after obtaining a VHG from the Wahnstr\"om (binary Lennard-Jones) glass.}
    \label{fig:Figure1}
\end{figure}

Virial homogenized glasses (VHG) are produced using an iterative scheme as illustrated in Fig.\ref{fig:Figure1}(a). We compute the local excess pressure contributions $P_{2,i}$ of particles in a glass, where $i$ is a particle index. If $P_{2,i}$ is larger (smaller) than the mean pressure ($\overline{P_{2,i}}$), then the particle is resized by a set amount such that the pressure is reduced (increased). Whether this is an increase or decrease in particle size will depend on the derivative of the pressure with respect to $\sigma_i$. If $dP_{2,i}/d\sigma_i > 0$, then $\sigma_i$ becomes $\sigma_i + \Delta\sigma$ if $P_{2,i} < \overline{P_{2,i}}$, and vice versa; if $dP_{2,i}/d\sigma_i < 0$, then the particle size corrections are reversed. The newly obtained configuration is then energy minimized using the FIRE algorithm \cite{Bitzek2006}. If the newly obtained configuration has a decreased standard deviation in virial pressure over the whole system $\Delta{P}_{2,i}$, then the  move is accepted and the process is repeated. If not, $\Delta\sigma$ is halved and the process repeated from the previous configuration, much like the variable step size of an optimization scheme. Note that $\overline{P_{2,i}}$ is recalculated after every newly accepted configuration and set of particle sizes, and the system volume is constant. The configuration is considered `converged' when five size change reductions have been carried out, but further reduction is needed to produce a configuration with a sharper $P_{2,i}$ distribution. We verified that considering more than 5 attempts to reduce $\Delta\sigma$ does not lead to significant further reductions in $\Delta{P_{2,i}}$.

The action of the algorithm can be traced by following distributions over iterations. For example, the distribution of $P_{2,i}$ is seen to sharpen significantly, as shown in Fig.\ref{fig:Figure1}(b); the standard deviation in $P_{2,i}$ is reduced by an order of magnitude by the end. The final size distribution is significantly wider than the original binary configuration, as shown in Fig.\ref{fig:Figure1}(c), but it is still possible to tell the two species apart. We also confirm that despite some modification, the local Voronoi volume fraction $\phi_i$ still has two peaks, as shown in SI Appendix. 

\subsection*{Simulations}
We consider a smoothed version (WAHNs) of the equimolar additive Lennard-Jones binary mixture introduced by Wahnstr\"om (WAHN) \cite{Wahnstrom1991} (see SI Appendix). This potential has been used to model metallic glasses and represents a good model system to monitor and correlate microscopic structural properties with the thermodynamics of the material. For example, below the glass transition temperature, $T_g$, icosahedra, shown by several works to be the locally favored structures (LFS) in this system \cite{Jenkinson2017,Wahnstrom1991,Malins2013,Coslovich2007,Pedersen2010,Pedersen2021}, are an indication of the stability of the glassy phase \cite{leoni2023a}, while above $T_g$, Frank-Kasper bonds (see \cite{Pedersen2010,leoni2023b}), are formed during crystallization \cite{leoni2023b}.

All simulations are run with the LAMMPS molecular dynamics package~\cite{LAMMPS} at constant volume. In the following, we report results in reduced units (energy in unit of $\epsilon$, distances in unit of $\sigma$, temperature in unit of $\epsilon/k_B$, see SI Appendix).

\section{Results and Discussion}\label{sec:results}

\subsection{Stability}
\begin{figure}[!t]
    \centering
        \includegraphics[width=0.99\linewidth]{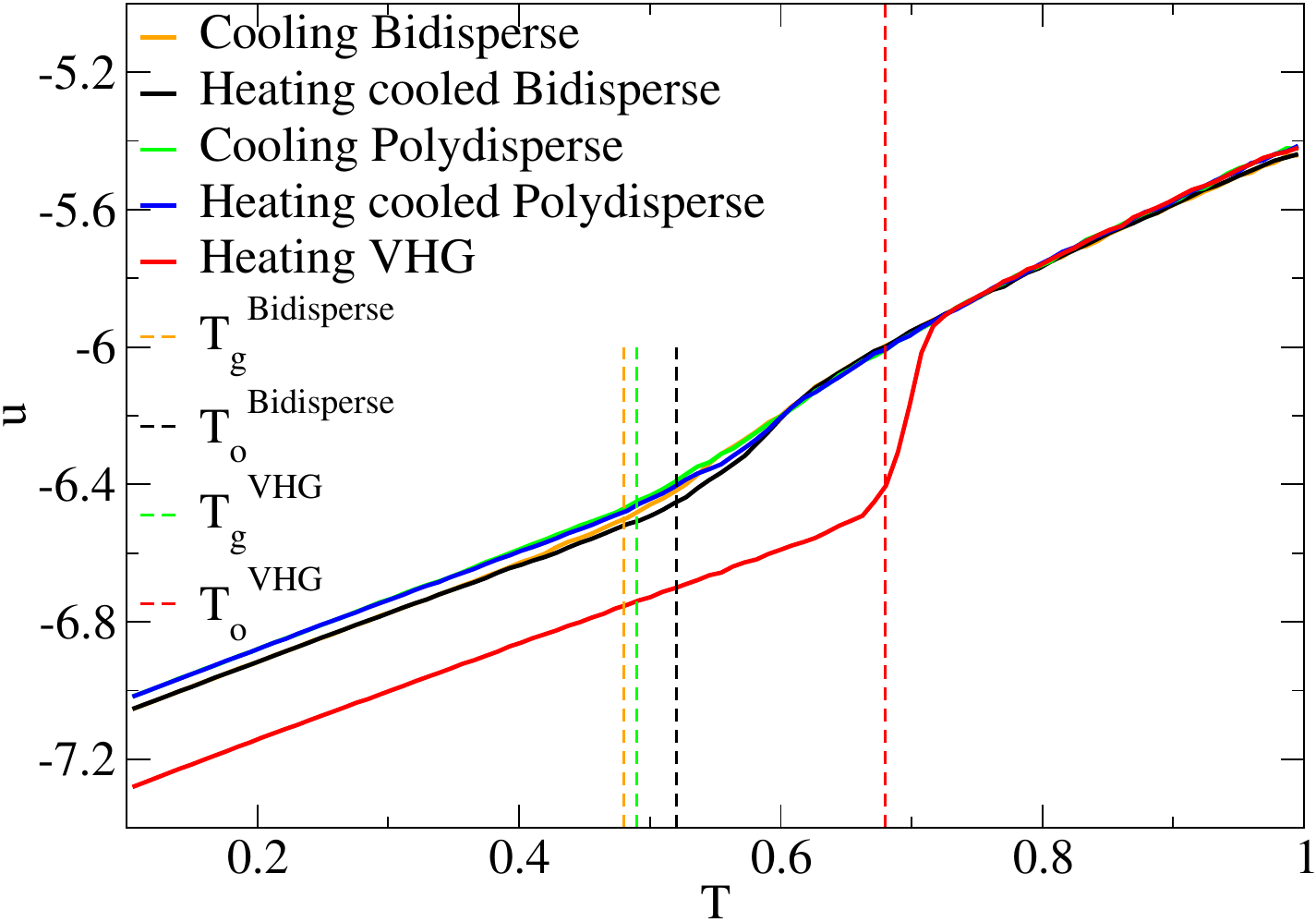}
    \caption{Potential energy per particle vs T for the bidpserse, polydisperse, and VHG systems. Vertical dashed lines indicate the estimation of $T_g$ and $T_o$ (see text). Note that the orange line almost coincides with the black line.} 
    \label{fig:Figure2}
\end{figure}

We start by cooling a bidisperse smoothed Wahnstr\"om mixture from $T=1.0$ to $T=0.1$ with a cooling rate of $\gamma=\Delta T/\Delta{t} =1.8\cdot 10^{-7}$ corresponding to a change of $\Delta T=0.9$ in $10^9$ integration steps (being $dt=0.005$ in reduced units). The potential energy during the cooling is plotted as an orange curve in Fig.~\ref{fig:Figure2}, where a supercooled liquid state is observed for $T\gtrsim 0.55$ followed by an out-of-equilibrium glassy state at $T\lesssim 0.55$, which is observed to have a linear dependence on temperature with a slope approaching the value $3k_BT/2$ at small $T$ (meaning that the state is trapped in an essentially harmonic well).

At the end of the cooling ramp we take the last configuration and apply the homogenization algorithm with an initial size adjustment factor $\Delta\sigma$ of 0.01, producing the VHG state. We then test the kinetic stability of both these configurations (before and after the homogenization algorithm) by heating them with a heating rate of $\gamma$. The heating curves are reported in Fig.~\ref{fig:Figure2} with a black and red line for the configuration before (bidisperse) and after (VHG) the homogenization algorithm. We can see very little difference in energy as a function of temperature between cooling and heating the original bidisperse mixture (orange and black curves, respectively). In contrast, we see that the VHG state (red curve) has increased considerably both its thermodynamic and kinetic stability: the former from the big drop in potential energy after the homogenization procedure (see Fig.~\ref{fig:Figure2}), the latter from the big increase in the onset or melting temperature of the state (see below for the specific values).

The homogenization procedure transforms the bidisperse system into a polydisperse one, though peaks corresponding to each species are still distinguishable (see curve {\it All} in Fig.~\ref{fig:Figure1}c). To exclude the possibility that the exceptional stability of the VHG glass is due to its polydispersity, we first melt the VHG glass to $T=1$ and then produce an annealed glass with the same cooling rate $\gamma$ as for the bidisperse system. The cooling of this polydisperse system is shown as the green curve in Fig.~\ref{fig:Figure2}, and is seen to mirror almost exactly the behaviour of the original bidisperse glass. This confirms that the stability of the VHG state is not associated with its size distribution, but rather from the fact that it is lying in a very deep minima of the potential energy landscape, which is not otherwise accessible via standard annealing.

\begin{figure}[!t]
    \centering
        \includegraphics[width=0.99\columnwidth]{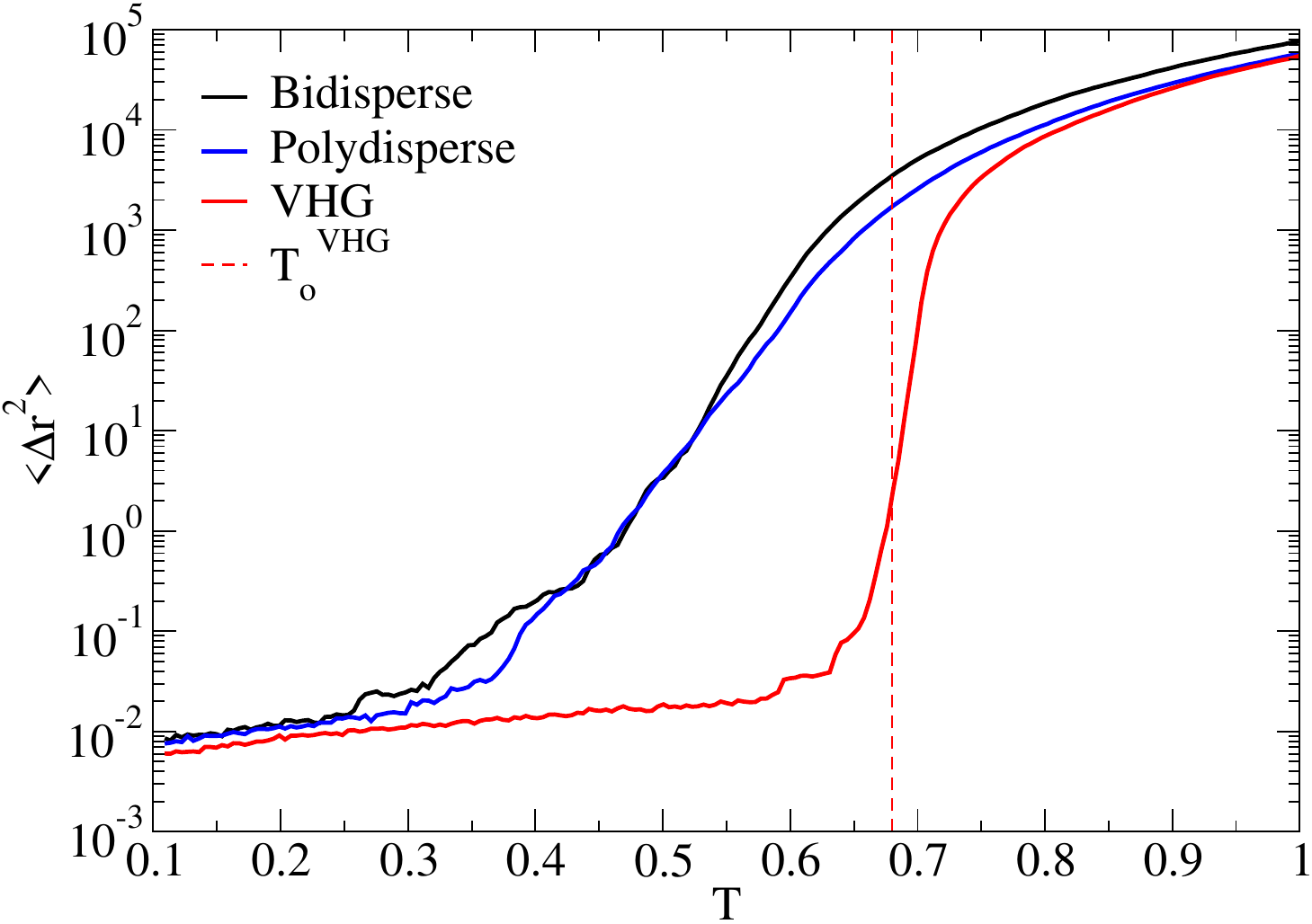}
    \caption{Comparison of heating ramps between bidisperse, polydisperse (obtained by melting and cooling VHG), and VHG glasses. $\langle\Delta r^2\rangle$ is the squared displacement of particles from their positions at $T = 0.1$ during each ramp.}
    \label{fig:Figure3}
\end{figure}

In Fig.~\ref{fig:Figure3}, we consider the dynamics of the three different systems (bidisperse, polydisperse, and VHG) during the heating ramps by plotting the mean square displacement $\langle\Delta r^2\rangle = 1/N\sum_{i=1}^{N}[r_i(T)-r_i(T=0.1)]^2$ of all N particles compared to their starting state at $T=0.1$. In agreement with the results obtained in Fig.~\ref{fig:Figure2}, we observe that the glass melting of the VHG system is more sharp and occurs at considerably higher temperatures compared to the bidisperse and polydisperse systems.

From the intersection between the glass line and the supercooled line shown by the potential energy per particle (Fig.~\ref{fig:Figure2}) we estimate for each model the glass transition temperature $T_g$ at the simulated cooling rate $\gamma$ and the onset or melting temperature $T_o$ at the simulated heating rate $\gamma$. We obtain the glass transition temperatures $T_g^{\rm Bidisperse}=0.48$ and $T_g^{\rm VHG}=0.49$, and the onset temperatures $T_o^{\rm Bidisperse}=0.52$ and $T_o^{\rm VHG}=0.68$, where the latter value corresponds to the temperature at which the mean squared displacement $\langle\Delta r^2\rangle$ versus T (Fig.~\ref{fig:Figure3}) upon heating undergoes a sudden increase (specifically, $T_o^{\rm VHG}=0.68$ corresponds to T($\langle\Delta r^2\rangle\simeq 1$)). We consider how far removed the onset temperature is with respect to the glass transition point, a common metric for characterizing ultrastable states \cite{Ediger2017,Rodriguez-Tinoco2022}: we obtain $T_o^{\rm VHG}/T_g^{\rm VHG}=0.68/0.49=1.39$. This is significantly larger with respect to what is usually obtained with vapor deposition, even though we expect this ratio to decrease for decreasing rate $\gamma$ going towards values comparable to typical experimental ones. For reference, the typical value obtained for vapor deposited organic molecules is $T_o/T_g\sim 1.05$ \cite{Ediger2017}.

\subsection{Locally favored structures}

% FIG 7
\begin{figure}[!t]
    \centering
        \includegraphics[width=0.99\columnwidth]{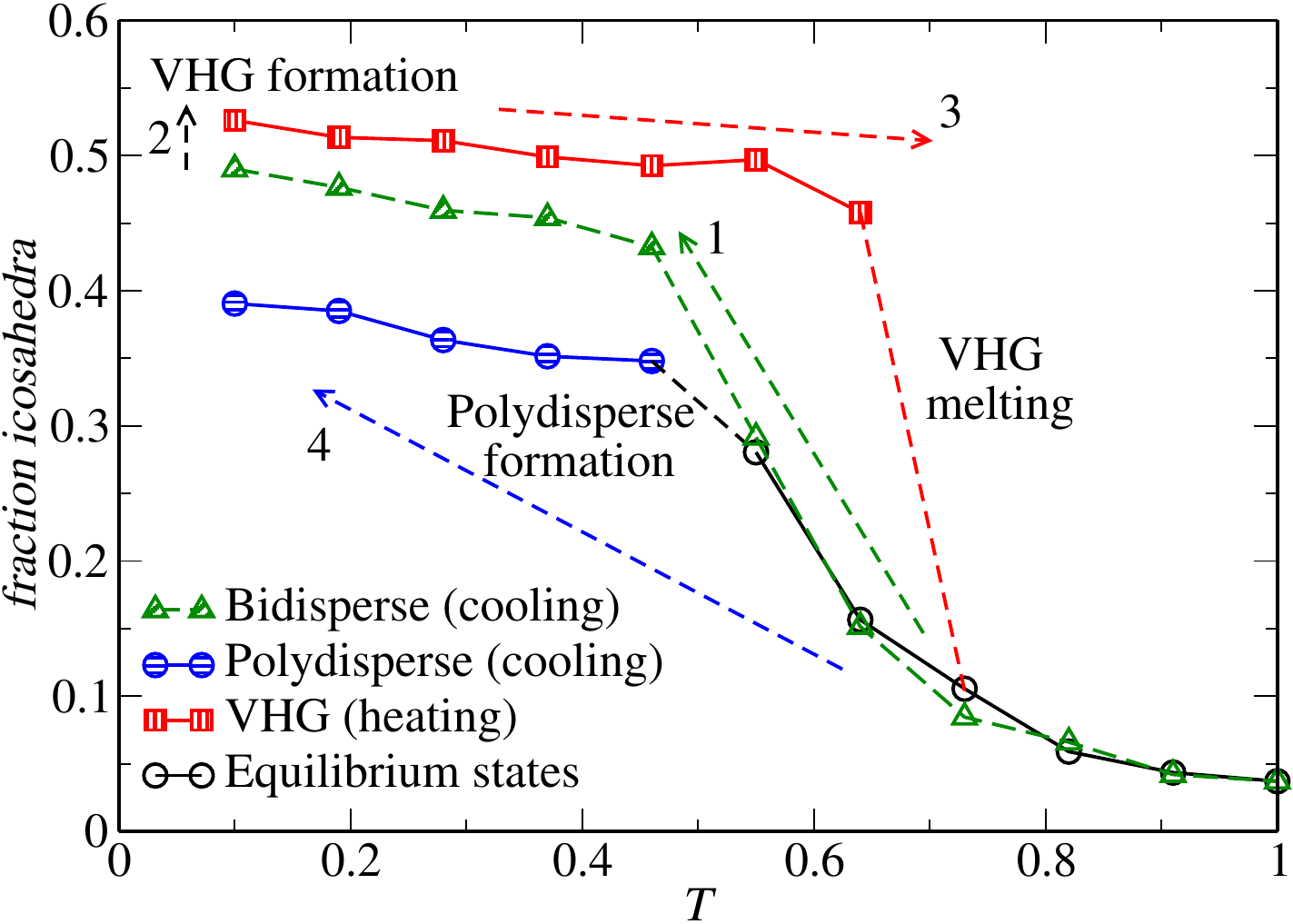}
    \caption{Fraction of icosahedra vs T for the following systems: cooling of the bidisperse system (green) and the polydisperse melt (blue), heating of VHG (red), and the liquid at equilibrium (black).}
    \label{fig:Figure4}
\end{figure}
%

% LFS: fraction of icosahedra
Both the fragility and the dynamic heterogeneities of the Wahnstr\"om system have been linked with an increase in the number of icosahedral environments that form around the small particles~\cite{Jenkinson2017,Wahnstrom1991,Malins2013,Coslovich2007,Pedersen2010,Pedersen2021}. In the glass phase both the composition and relative orientation of the icosahedra are disordered and distinct from those found in the crystalline structure (the Laves phase $A_2B$~\cite{Pedersen2010}). A similar correlation has been recently found in a vapor deposited ultrastable glass of Wahnstr\"om particles \cite{leoni2023a}, where the fraction of icosahedral environments was linked with its stability. We thus study the population of icosahedral environments in VHG glasses, identifying icosahedra with the Voronoi topology based tool VoroTop \cite{lazar2018}. Fig.~\ref{fig:Figure4} shows the fraction of icosahedra with respect to the small particles (which in the original Wahnstr\"om model have diameter $\sigma_{11}$ and make up half of the total) versus temperature. We consider the cooling of the original bidisperse system, the VHG state during melting, and cooling of the same polydisperse particle population all performed at the same rate $\gamma$. Characterisation of equilibrium liquid configurations are also presented for comparison. While increasing polydispersity usually disrupts icosahedral environments, the first surprising result is that the homogenization procedure produces a higher fraction of icosahedral environments, despite the increased polydispersity. If we compare the polydisperse system and the VHG state, which have the same size distribution, we see that the VHG state is considerably richer in icosahedral clusters. These results confirm that there is a strong link between glass stability and the fraction of icosahedral environments, and that VHG states are thermodynamically, kinetically, and structurally more stable.

\subsection{Relaxation dynamics}

\begin{figure}[!t]
    \centering
        \includegraphics[width=1\columnwidth]{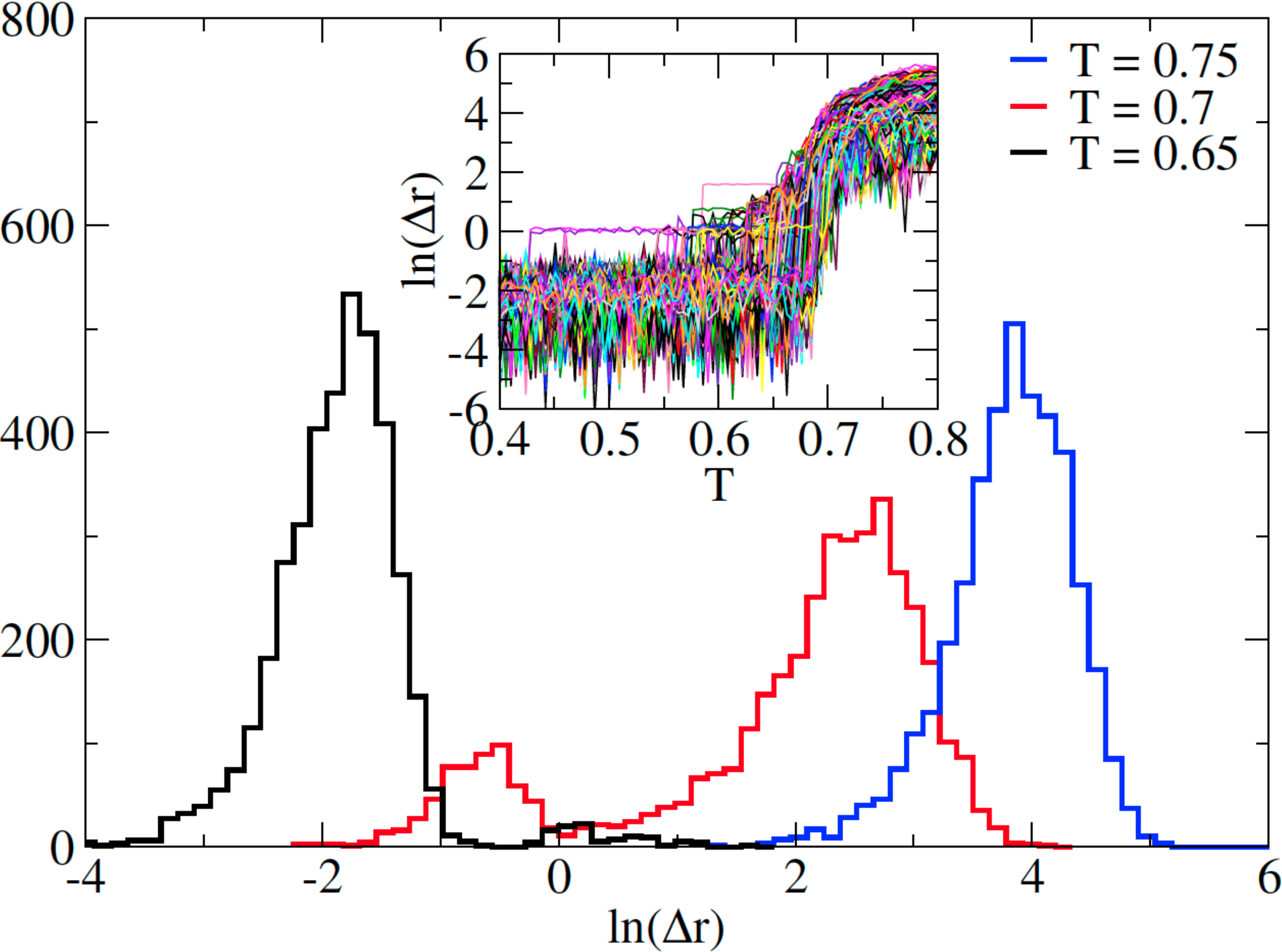}
    \caption{Histogram of the logarithm of the displacements during the heating ramp of the VHG system at three specific temperatures showing the presence of two populations (at $T=0.65$ and $T=0.7$), one of caged (left peak) and the other of diffusing (right peak) particles. At $T=0.75$ no more particles are caged. Inset: Logarithm of the displacement vs T for all particles during the heating ramp (here in the range $T=0.4-0.8$, full range in SI Appendix) applied to the VHG system. Around $T=0.6$ diffusing particles (with $\Delta r>1$) start to appear. These particles are responsible for the appearance of avalanches resulting in a stepped  curve in the range $T=0.6-0.7$ in Fig.~\ref{fig:Figure3}.}
    \label{fig:Figure5}
\end{figure}

% Dynamics
From the curve of the mean squared displacement of the VHG glass in Fig.~\ref{fig:Figure3}, we notice that during heating, the system moves from a frozen to a diffusive regime through a step-like dynamics around temperatures in the range $0.6\leq T\leq 0.65$. This suggests the presence of avalanche-like behavior during glass melting~\cite{Sanz2014,Yanagishima2017}. In Fig.~\ref{fig:Figure5} we plot the logarithm of the displacement of all particles in the VHG state during the heating ramp, where we observe the presence of a population of particles which exhibit occasional jumps before the melting temperature, on the order of a particle diameter. In Fig.~\ref{fig:Figure5} we plot the histogram of the logarithm of displacements measured at three different temperatures. At $T=0.65$ (in black) two populations of particles can be distinguished based on their displacement $\Delta r$ from the initial configuration: a large population with small $\Delta r$ forming the left-peak corresponding to the particles who are trapped in their cages and exhibit only vibrational motion, and a (very) small population of particles that undergo displacement comparable to the particle size ($\Delta r \approx 1)$. At $T=0.7>T_g^{\rm VHG}$ (in red) the fast peak already comprises the majority of particles, while at $T=0.75$ (in blue) all particles have diffused several particle diameters.

\begin{figure}[!t]
    \centering
        \includegraphics[width=0.49\columnwidth]{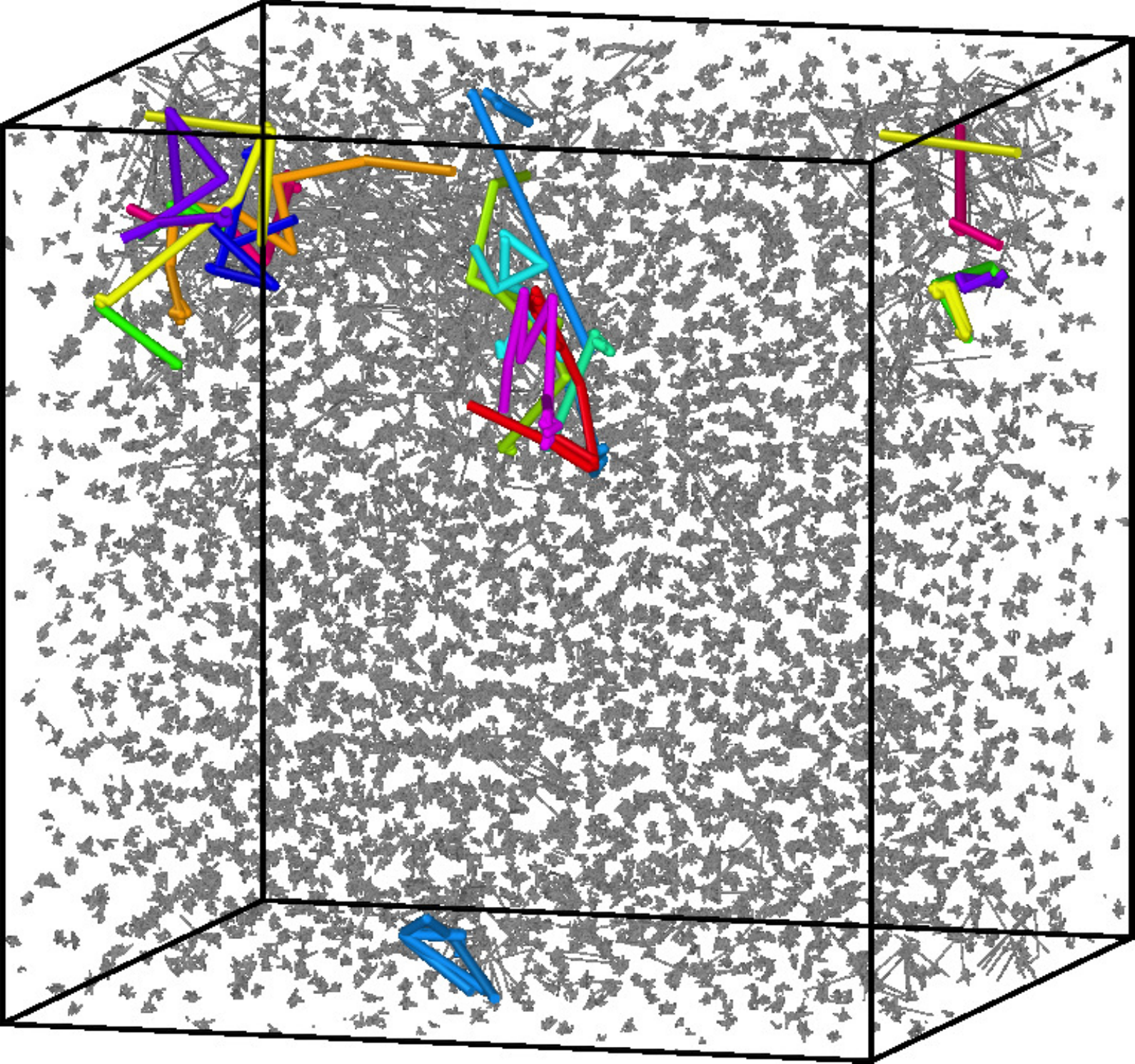}
        \includegraphics[width=0.49\columnwidth]{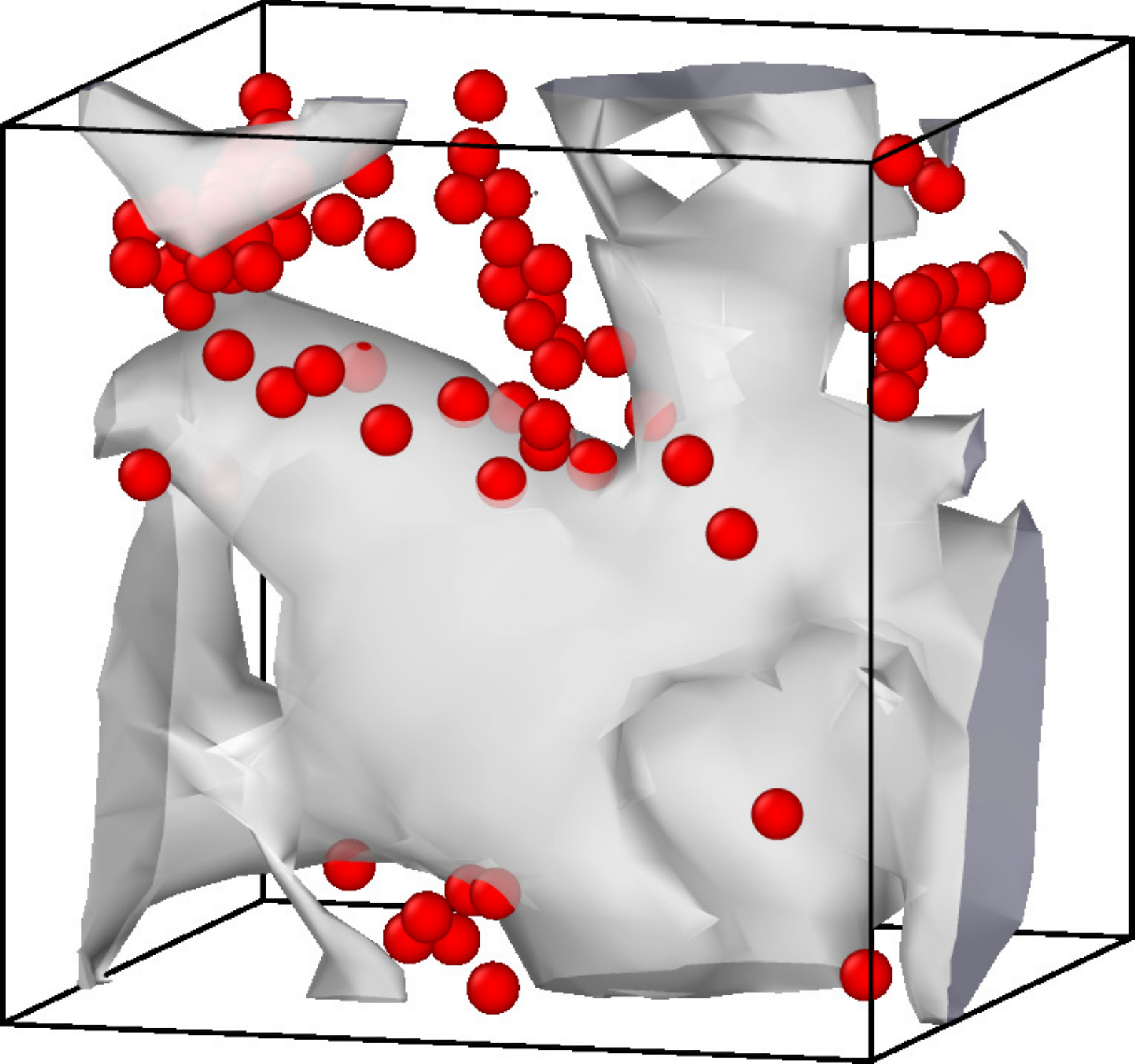}
    \caption{Left: trajectories of all particles of the VHG system during the heating ramp taken from $T=0.6$ to $T=0.675$. Gray trajectories show particles whose displacement at $T=0.65$ is less than 2 starting from the $T=0.6$ state. The trajectories of particles whose displacement is greater than 2 (14 trajectories in total) are plotted in colour. Right: Slice showing particles with displacement variation $\delta^*>0.3$. The shaded region is obtained from a spatial mesh of particles identified as icosahedra at $T=0.65$.}
    \label{fig:Figure6}
\end{figure}

These results show clearly that the melting of the bulk ultra-stable glass occurs via a well-defined population of particles with intermittent (avalanche-like) dynamics. In the left panel of Fig.~\ref{fig:Figure6} we plot the trajectories of all particles during the heat ramp for the temperature range going from $T=0.6$ to $T=0.675$: we colour in gray the trajectories of the particles that have moved less than 2 particle diameters, and in color the 14 trajectories of the particles that have moved more then 2 particle diameters. From the figure it is clear that the fast particles are spatially correlated and appear in compact regions. These results confirm the presence of localized fast regions~\cite{ruiz2023real} and that glass melting is akin to a nucleation-and-growth process~\cite{vila2023emergence}.

By computing the variation of the displacement $\delta^*$ (see SI Appendix) undergone by each particle in the range of temperatures $T=0.6-0.65$ where the system changes dynamical regime, we can distinguish a subset of particles displacing at larger distances with respect to the rest of the system. The value of $\delta^*$ separating these two subset of particles is conveniently fixed to 0.3 (see SI Appendix). The subset of particles experiencing the largest displacement variation is large enough (129 particles) to make an histogram of the size $\sigma$ of the particles. This histogram shows a clear correlation between larger displacement and small particle size (see Fig.S2 in SI Appendix). Furthermore, from the right panel of Fig.~\ref{fig:Figure6} we can see a slice of the box including particles with $\delta^*>0.3$ which are found in regions of the system mainly free from icosahedra formation (indicated with a grey shaded region). This shows that glass melting starts in the regions devoid of locally favoured structures.

\section{DISCUSSIONS AND CONCLUSIONS}
\label{sec:conclusions}

We briefly consider how numerical studies like this might inform the experimental realization of ultra-stable colloidal packings. Though evidence of suppressed particle dynamics has been reported for two-dimensional systems \cite{Cao2017}, the realization of bulk ultra-stable colloidal glasses remains difficult. The single particle editing required for {\it swap} or the VHG protocol would be impractical. The random bonding suggested by Ozawa {\it et al.}~\cite{ozawa2023creating} could be realized using particles with directional interactions, but it is unclear whether this would be a route of choice for arbitrary particle populations However, even though particle resizing itself is potentially prohibitive, our work suggests that the homogenization of mechanical environments might be an effective route to improving the stability of glasses. This would require an appropriate annealing procedure. Despite there being no reports of ultrastability through annealing alone as of yet, mechanically or thermally annealed states do show some of the hallmarks of a more equilibrated glass, such as brittle fracture \cite{Leishangthem2017,Bhaumik2021}, and reduced energy/enhanced modulus \cite{Priezjev2019}. We note that heterogeneity in local elasticity in metallic glasses, most often expressed as a heterogeneous local yield stress, has been shown to lead to material softening (as reviewed in \cite{Egami2013}): this would suggest that their removal would yield the hardest, most equilibrated glass. The feasibility of applying a novel annealing procedure to realizing the emergence of mechanical homogeneity and ultrastability remains a topic for future work.

To summarize, we explored a new protocol for the preparation of ultrastable glasses using an algorithm that homogenizes local virial stress fluctuations through incremental changes to individual particle sizes. Applying this algorithm to an additive Lennard-Jones mixture, we observed significant improvements in both the thermodynamic, kinetic and structural stability of the glass. Thermodynamic stability is seen in a sharp decrease in the potential energy of the system. Kinetic stability was observed in a considerable increase in the onset or melting temperature of the glass and in the transition becoming more sharp. Structural stability was inferred by a strong increase in locally favoured structures, which for this system are the icosahedral environments. We then considered the melting behaviour of the ultra-stable glass, finding that bulk melting starts from localized regions devoid of icosahedral ordering. The motion in these regions is characterized by avalanche-like intermittent motion of a small number of particle with a lower than average radius, which then trigger the melting of the rest of the glass.

Our results provide strong evidence for a deep link between thermodynamic and mechanical stability of ultra-stable glassy states. The observed virial stress homogenization, in addition to the well-established link between stability and the fraction of icosahedral structures, provides valuable insights into the mechanisms governing glass stabilization. In this respect, the regions with virial stress heterogeneity can be seen as mechanical defects whose suppression considerably increases the thermodynamic stability of the ultra-stable state, i.e. the height of the energy barriers that separate it from nearby minima.

\vspace{0.5cm}

\section{Acknowledgments}

\begin{acknowledgments}
F.L. and J.R. acknowledge support from the European Research Council Grant DLV-759187. F.L., J.R. and F.S. acknowledge support by ICSC – Centro Nazionale di Ricerca in High Performance Computing, Big Data and Quantum Computing, funded by European Union – NextGenerationEU, and CINECA-ISCRA for HPC resources. T.Y. acknowledges the Kyoto University Foundation and support through JST CREST Grant Number JPMJCR2095.
\end{acknowledgments}

%\newpage
\bibliography{ref}

\clearpage
\onecolumngrid

\begin{center}
{\bf\large{Supplementary Material for ``Generating ultrastable glasses by homogenizing the local virial stress''}}
\end{center}

\setcounter{equation}{0}
\setcounter{figure}{0}
\setcounter{table}{0}
\setcounter{section}{0}
\makeatletter
\renewcommand{\theequation}{S\arabic{equation}}
\renewcommand{\thefigure}{S\arabic{figure}}
\renewcommand{\thetable}{S\arabic{table}}
\renewcommand{\thesection}{S\arabic{section}}

\section{Local volume fraction distribution in VHG state}
\begin{figure}[!h]
    \centering
        \includegraphics[width=0.5\columnwidth]{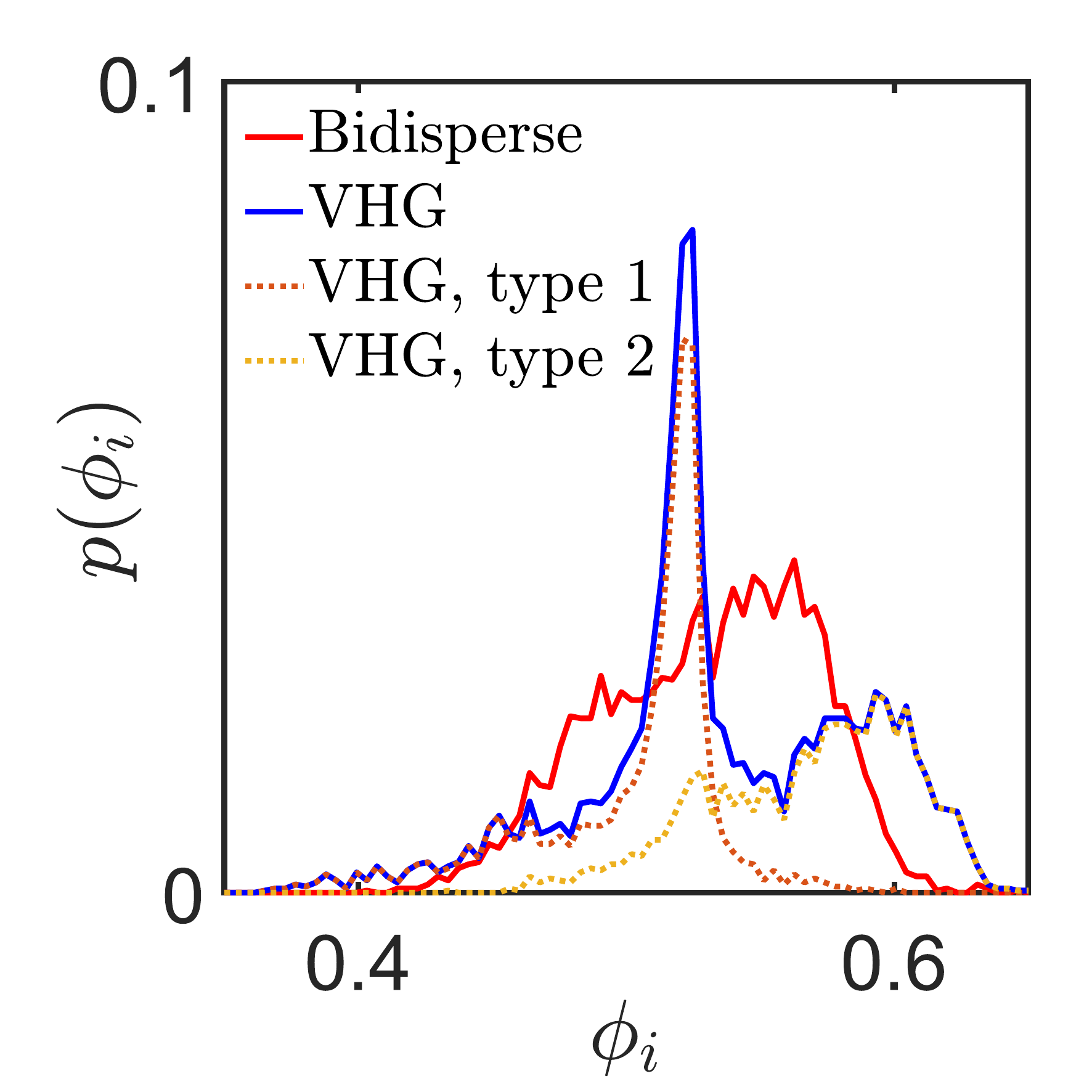}
    \caption{Distribution of local volume fractions in the VHG state.}
    \label{fig:SFig1}
\end{figure}

In previous work~\cite{Yanagishima2021}, it was found that homogenization of local volume fractions through local particle size adjustments to a monodisperse Weeks-Chandler-Andersen glass could realize enhanced stability against spontaneous devitrification. However, as noted in the main text, this is only effective in the context of a single-component system. More generally, in order to homogenize local mechanical environments, the strategy introduced in this work is more effective. The volume fraction distribution of the originally two component VHG state discussed in the main text is given in Fig.~\ref{fig:SFig1}, calculated using individual particle sizes $\sigma_i$ and individual Voronoi volumes using a radical Voronoi tessellation, $\phi_i=\pi\sigma_i^3/6$. It is clear that the VHG algorithm does not cause a collapse of $\phi_i$. While distributions for both particle types are wider, the two species are still distinguishable.

\newpage

\section{Smoothed Wahnstr\"om (WAHNs) potential model}

The original WAHN model \cite{Wahnstrom1991} is described by the equimolar additive Lennard-Jones (LJ) binary mixture potential
\begin{equation}
u_{LJ}(r)=4\epsilon_{\alpha\beta}\left[\left(\dfrac{\sigma_{\alpha\beta}}{r}\right)^{12}-\left(\dfrac{\sigma_{\alpha\beta}}{r}\right)^6\right],
\end{equation}
where $\alpha$ and $\beta$ can be particles of type 1 and 2. The energies are $\epsilon_{11}=\epsilon_{12}=\epsilon_{22}=\epsilon$, the masses $m_2/m_1=2$, the diameters $\sigma_{22}/\sigma_{11}=1.2$, and the cross-interaction diameter given by the Lorentz-Berthelot mixing rule: $\sigma_{12}=(\sigma_{11}+\sigma_{22})/2$. The unit of length is set to $\sigma=sigma_{11}$. We apply two changes to the original WAHN model: one is to consider the ratio between the masses proportional to the ratio of the respective spheres volume (i.e., $m_2/m_1=(\sigma_{22}/\sigma_{11})^3=1.728$) to have a criteria to assign the mass to each particle in the case of the polydisperse system. The other change consists on smoothing the potential with a polynomial function such that at the outer cutoff ($r_{out}$), the potential and its first (i.e., the force) and second derivatives go continuously to zero. The resulting potential in the WAHNs model, $u_s$, is described by the following equations:
\begin{equation}
\begin{array}{llll}
u_s & = & 4\epsilon\left[\left(\dfrac{\sigma}{r}\right)^{12}-\left(\dfrac{\sigma}{r}\right)^6\right]-u_{o} & \ \ \ \ 0<r<r_{in}\\
&&\\
u_s & = & C_0-C_1(r-r_{in})-\dfrac{C_2}{2}(r-r_{in})^2-\dfrac{C_3}{3}(r-r_{in})^3-\dfrac{C_4}{4}(r-r_{in})^4 & \ \ \ \ r_{in}<r<r_{out}\\
&&\\
u_s & = & 0 & \ \ \ \ r>r_{out},
\end{array}
\end{equation}
where we choose the inner and outer cutoff $r_{in}=2.5\sigma_{ij}$ and $r_{out}=2.75\sigma_{ij}$, respectively. The 3 continuity conditions at $r=r_{in}$ and the 3 conditions at $r=r_{out}$ (going continuously to zero) for the potential and its first and second derivatives give the following values for the 6 parameters describing $u_s$:
\begin{equation}
\begin{array}{lll}
C_1 & = & 24\dfrac{\epsilon}{r_{in}}\left[2\left(\dfrac{\sigma_{ij}}{r_{in}}\right)^{12}-\left(\dfrac{\sigma_{ij}}{r_{in}}\right)^6\right] \\
&&\\
C_2 & = &  -24\dfrac{\epsilon}{r_{in}^2}\left[26\left(\dfrac{\sigma_{ij}}{r_{in}}\right)^{12}-7\left(\dfrac{\sigma_{ij}}{r_{in}}\right)^6\right]\\
&&\\
C_3 & = & -\dfrac{3}{(r_{out}-r_{in})^2}\left[C_1+\dfrac{2}{3}C_2(r_{out}-r_{in})\right] \\
&&\\
C_4 & = & -\dfrac{1}{3(r_{out}-r_{in})^2}\left[C_2+2C_3(r_{out}-r_{in})\right] \\
&&\\
C_0 & = &  C_1(r_{out}-r_{in})+\dfrac{C_2}{2}(r_{out}-r_{in})^2+\dfrac{C_3}{3}(r_{out}-r_{in})^3+\dfrac{C_4}{4}(r_{out}-r_{in})^4\\
&&\\
u_{o} & = & 4\epsilon\left[\left(\dfrac{\sigma_{ij}}{r_{in}}\right)^{12}-\left(\dfrac{\sigma_{ij}}{r_{in}}\right)^6\right]-C_0     
\end{array}    
\end{equation}
The smoothing of the LJ potential described in this section is implemented in LAMMPS \cite{LAMMPS} through the commands: \textit{pair\_style lj/smooth} in combination with \textit{pair\_modify shift}.  
In the main text we consider $N=4000$ particles in total at the reduced number density $\rho^*=\rho\sigma^3=0.75$ which has been considered in the original work by Wahnstr\"om \cite{Wahnstrom1991} and in later works \cite{leoni2023a,leoni2023b}.

\clearpage
\newpage

\section{Relaxation dynamics}

\begin{figure}[!h]
    \centering
        \includegraphics[width=0.5\columnwidth]{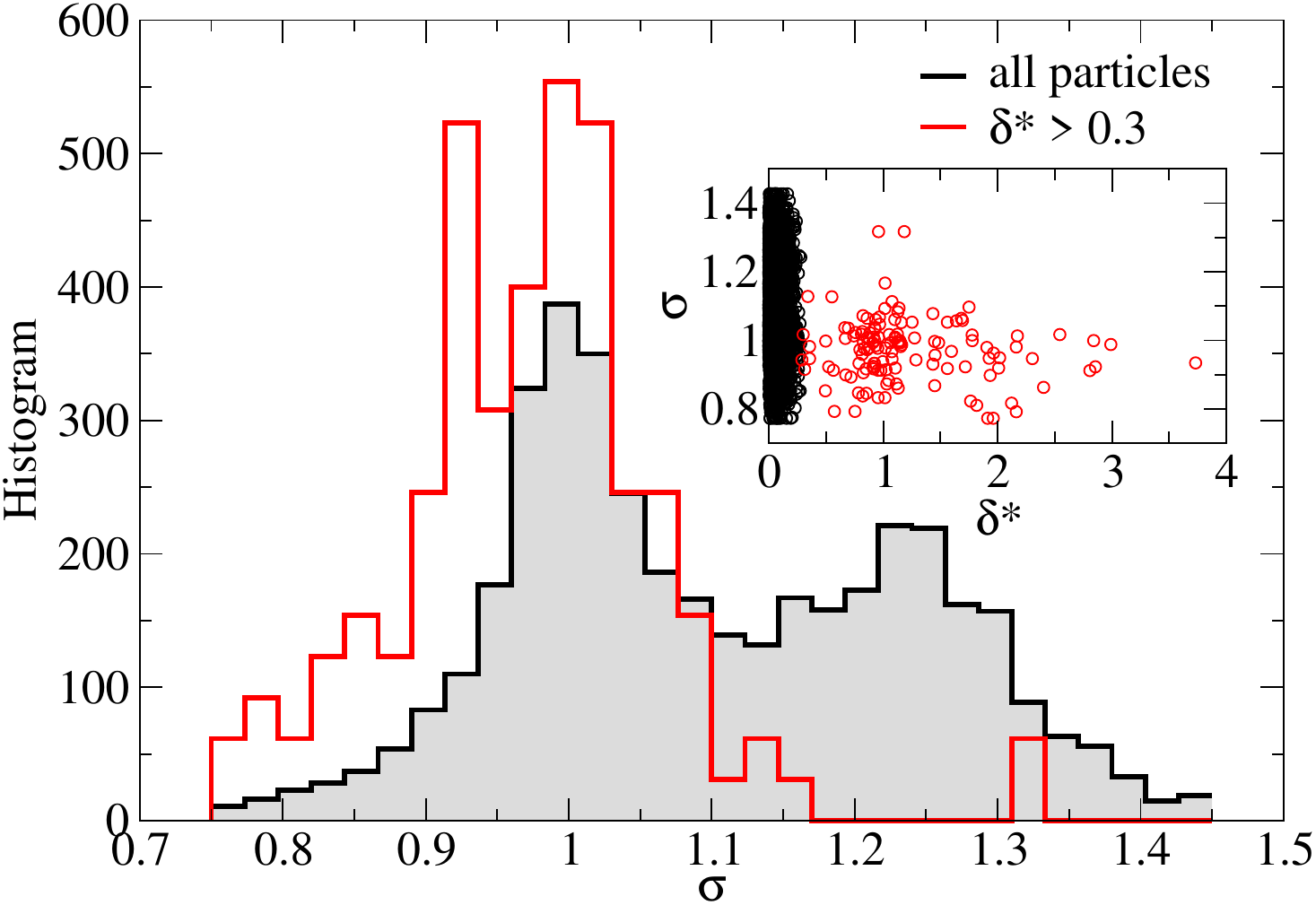}
    \caption{Histogram of particle diameters $\sigma$ for the full system (black) compared to the normalized histogram of $\sigma$ for particles with $\delta^*>0.3$ (red). Inset: $\sigma$ versus $\delta^*$ of all particles of the system for $\delta^*<0.3$ (black) and $\delta^*>0.3$ (red).}
    \label{fig:SFig2}
\end{figure}

In the range of temperatures $T=0.6-0.65$ where the system changes dynamical regime, it is possible to distinguish a subset of particles displaced by larger distances with respect to the rest of the system. This subset is defined by particles with $\delta^*>0.3$, where $\delta^*=|\langle\Delta r(T=0.65)\rangle-\langle\Delta r(T=0.6)\rangle|$, with $\langle\Delta r(T)\rangle=\sum_{i=1}^{N}\sqrt{(x_i(T)-x_i(0.1))^2+(y_i(T)-y_i(0.1))^2+(z_i(T)-z_i(0.1))^2}/N$. The inset of Fig.~\ref{fig:SFig2} shows the value of $\sigma(\delta^*)$ for every particle. Red circles correspond to the 129 particles with $\delta^*>0.3$. In Fig.~\ref{fig:SFig2}, we can see the normalized histograms (integrals matched to $N=4000$ for both the subset and the full system).

\section{Displacement}

\begin{figure}[!h]
    \centering
        \includegraphics[width=0.5\columnwidth]{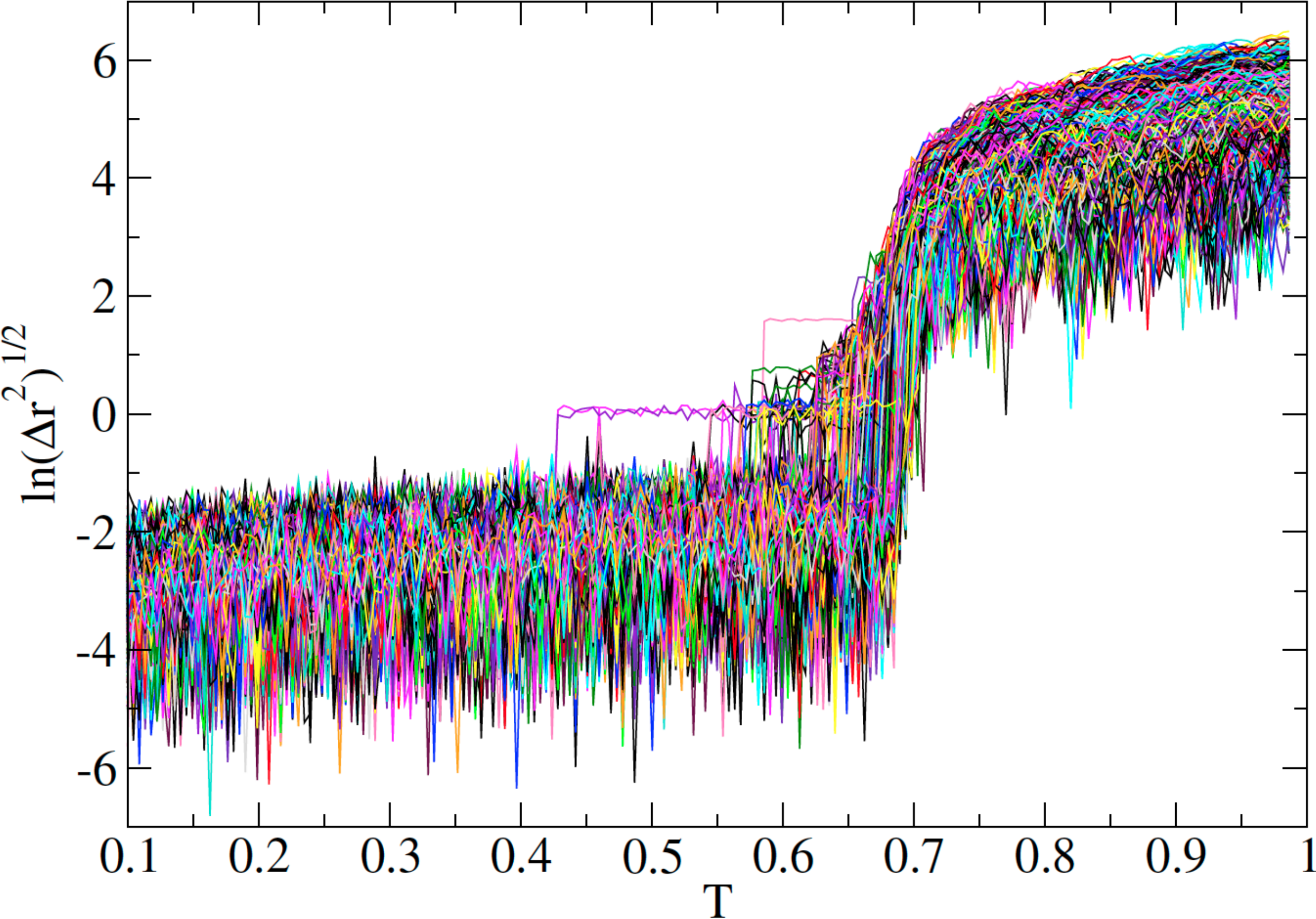}
    \caption{Logarithm of the displacement vs T for all particles during the heating ramp applied to the VHG system.}
    \label{fig:SFig3}
\end{figure}

\end{document}